\documentclass{JHEP3} 

\usepackage{epsfig,multicol}
\usepackage{amsmath, amssymb}

\newcommand\fverb{\setbox\pippobox=\hbox\bgroup\verb}
\newcommand\fverbdo{\egroup\medskip\noindent%
			\fbox{\unhbox\pippobox}\ }
\newcommand\fverbit{\egroup\item[\fbox{\unhbox\pippobox}]}
\newbox\pippobox

\newcommand{\be}{\begin{equation}} 
\newcommand{\ee}{\end{equation}}

\newcommand{\ba}{\begin{eqnarray}}
\newcommand{\ea}{\end{eqnarray}}

\newcommand{\zz}{\mathbb{Z}}
\newcommand{\rr}{\mathbb{R}}

\newcommand{\la}{\longrightarrow}

\newtheorem{theorem}{Theorem}[section]
\newcommand{\proof}[1]{{\em Proof:} {#1} \vskip 1pt\noindent{\hfill{\it $\Box$\hskip 1cm}}}

\newcommand{\ads}{AdS_5\times S^5}

\title{Highest states in light-cone $AdS_5\times S^5$ superstring}

\author{Matteo Beccaria\\
        Dipartimento di Fisica, Universita' di Lecce,
        Via Arnesano, 73100 Lecce\\
        INFN, Sezione di Lecce\\
        E-mail: \email{matteo.beccaria@le.infn.it}}

\author{Gian Fabrizio De Angelis\\
        Dipartimento di Fisica, Universita' di Lecce,
        Via Arnesano, 73100 Lecce\\
        INFN, Sezione di Lecce\\
        E-mail: \email{deangelis@le.infn.it}}

\author{Luigi Del Debbio \\
        SUPA, School of Physics, University of Edinburgh,
        Edinburgh EH9 3JZ, UK\\
        E-mail: \email{luigi.del.debbio@ed.ac.uk}
}

\author{Marco Picariello\\
        Dipartimento di Fisica, Universita' di Lecce,
        Via Arnesano, 73100 Lecce\\
        INFN, Sezione di Lecce\\
        E-mail: \email{marco.picariello@le.infn.it}}

\preprint{}

\abstract{ We study the highest states in the compact rank-1 sectors
of the $\ads$ superstring in the framework of the recently proposed
light cone Bethe Ansatz equations.  In the $\mathfrak{su}(1|1)$ sector
we present strong coupling expansions in the two limits $L,
\lambda\to\infty$ (expanding in power of $\lambda^{-1/4}$ with fixed
large $L$) and $\lambda, L\to\infty$ (expanding in power of $1/L$ with
fixed large $\lambda$) where $\lambda$ is the 't Hooft coupling and
$L$ is the number of Bethe momenta.  The two limits do not commute
apart from the leading term which reproduces the result obtained with
the Arutyunov-Frolov-Staudacher phase in the $\lambda, L\to\infty$
limit.  In the $\mathfrak{su}(2)$ sector we perform the strong coupling
expansions in the $L\to\infty$ limit up to ${\cal O}(\lambda^{-1/4})$,
and our result is in agreement with previuos String Bethe Ansatz analysis.}


\begin{document} 

\section{Introduction}
\label{Sec:Intro}

The verification of the conjectured AdS/CFT
correspondence~\cite{Maldacena:1997re,Gubser:1998bc,Witten:1998qj,Kazakov:2004qf,Klebanov:2000me}
is a non--trivial interpolation problem between a pair of quite
different theories, string theory on $AdS_5\times S^5$ and the
maximally supersymmetric four dimensional ${\cal N}=4$ super
Yang-Mills SU(N) gauge theory (SYM).  The duality predicts that
certain SYM gauge invariant operators have anomalous dimensions equal
to the energy of dual massive string states. Even though a drastic simplification
is achieved in the planar limit $N\to\infty$ with fixed 't Hooft
coupling $\lambda =g^2_{\rm YM}\,N$, both quantities can not be
simultaneously computed in the full range of the 't Hooft coupling.

On the gauge side, weak coupling perturbation theory provides explicit
results at a {\em relatively} low loop
order~\cite{ThreeLoops,Bern:2006ew}.  Assuming quantum integrability,
it is possible to identify the dilatation operator with an integrable
quantum Hamiltonian and write its Bethe Ansatz (BA)
equations~\cite{GaugeBA}.  They are conjectured to reproduce the full
weak coupling expansion of anomalous dimensions in various closed
sectors of the full $PSU(2,2|4)$ symmetry group. This is true up to
wrapping problems which occur at order ${\cal O}(\lambda^L)$ where $L$
is the classical dimension of the composite operator. This is an
essential limitation to the possibility of computing the all-order
weak coupling expansion of operators with fixed dimension.  Some
investigations suggest that wrapping problems could be overcome in the
fermionic approach based on the Hubbard
model~\cite{Rej:2005qt,Minahan:2006zi,Beccaria:2006aw,Beccaria:2006pb,Feverati:2006hh}.
However, even if wrapping problems could be solved, it would be
non--trivial to extrapolate the BA predictions to strong coupling. It
would also be necessary to prove that eventual non--perturbative
effects are captured by the gauge BA equations.

On the string side, the exact quantization of type IIB superstring on
$AdS_5\times S^5$ is not known. In the usual approach, one starts with
an exact supergravity solution at large $\lambda$ and computes
perturbative $\sigma$-model corrections.  The accessible regions in
the gauge and string theories are apparently disjoint.

An overlap window opens as soon as BMN-like scaling limits are
taken~\cite{Berenstein:2002jq}.  One considers classical solutions
with large angular momenta $J$ on $S^5$ (and/or spin on $AdS_5$ in
more general cases).  On the gauge side $J$ is the $R$-charge of the
dual composite operator. In the strict BMN limit, $J\to\infty$ with
fixed $\lambda' = \lambda/J^2$, or in the near-BMN corrections
suppressed by $1/J$ factors, the two calculations can be compared
because $\lambda$ is large while the gauge theory effective coupling
$\lambda'$ can be small. The comparison reveals a typical three loop
disagreement (see for instance~\cite{Minahan:2006sk} for a review).
Understanding the precise mechanism behind this discrepancy is a main
open problem in AdS/CFT.  It has been suggested that it arises because
the string and gauge calculations are performed in the double limit
$J\to\infty$ and $\lambda'\to 0$ taken in opposite
order~\cite{GaugeBA}.  This is an essential obstruction. To match
string calculations one should at least resum the weak coupling
perturbative series before taking the $J\to\infty$ limit, something
which is forbidden by the wrapping problems.  Unfortunately, we lack
the necessary technical tools to perform such resummations,

For these reasons, it is sensible to try to look for string BA
equations encoding the $\sigma$-model corrections.  At the classical
level, the $AdS_5\times S^5$ superstring~\cite{Metsaev:1998it} is
integrable~\cite{Bena:2003wd}. Assuming that the integrable structure
can be maintained at the quantum level, string BA equations (SBA) have
been proposed in~\cite{Arutyunov:2004vx}. They are similar in
structure to the gauge BA equations, but are modified by a non trivial
dressing phase~\cite{Beisert:2005tm,Beisert:2005wv}.

To fix the dressing phase, the SBA equations have been deeply tested
by comparing their predictions with the semiclassical quantization of
pp-wave states and spinning string solutions where available (see for
instance~\cite{OneLoopChecks}).  For pp-wave states, explicit string
theory calculations including curvature corrections to the flat space
background predict anomalous dimensions with the typical
form~\cite{Callan} 
\be 
\Delta^{\rm pp} -J = \Delta_0^{\rm
pp}(\lambda') + \frac{1}{J}\,\Delta_1^{\rm pp}(\lambda') + \cdots .
\ee
Both the thermodynamical limit $\Delta_0^{\rm pp}(\lambda')$
(independent on the dressing phase) and the first quantum correction
$\Delta_1^{\rm pp}(\lambda')$ are in full agreement with the explicit
string calculation.  For spinning string states the comparison is more
problematic. In this case, it is customary to introduce ${\cal J} =
J/\sqrt\lambda$ and the typical prediction for the semiclassical
energy is
\be
\Delta^{\rm FT} = \sqrt{\lambda}\, \Delta_0^{\rm FT}({\cal J}) + \Delta_1^{\rm FT}({\cal J}) + \cdots,
\ee
(with FT standing for Frolov-Tseytlin).  Now, the agreement with
explicit string calculations is perfect for $\Delta_0^{\rm FT}$ where
it holds by construction, but only partial in $\Delta_1^{\rm FT}$.
The problem has been recently clarified in
~\cite{Schafer-Nameki:2006ey} with an explicit comparison in the case
of the $\mathfrak{sl}(2)$ spinning string~\cite{Park:2005ji} and using
the full available information about the dressing
phase~\cite{Freyhult:2006vr,Arutyunov:2006iu,Janik:2006dc}.  At large
${\cal J}$ the exact string calculation admits the expansion
\be
\Delta^{\rm FT}_1({\cal J}) = \sum_{\ell\ge 2}\frac{f_\ell}{{\cal J}^\ell} + \sum_{s\ge 0} a_s\, e^{-2\pi\,s\, {\cal J}}.
\ee
The SBA equations are known to reproduce the full power series, but
not the exponentially suppressed terms.  The reason behind this
failure in capturing non-perturbative finite size corrections is a
fundamental limitation of any approach based on the thermodynamical
classical Bethe Ansatz.  It is not clear whether this problem could be
solved by resumming the conjectured all-order strong coupling series
for the dressing factor~\cite{Beisert:2006ib,Beisert:2006ez}.  The
resolution of this discrepancy could require the introduction of new
degrees of freedom in the quantum Bethe Ansatz as pointed out very
clearly in~\cite{Schafer-Nameki:2006ey}.

An outcome of the above discussion is that it is definitely very
important to test the SBA equations in all possible ways.  In this
spirit, apart from states admitting BMN-like limits, another important
structural test of the SBA equations is the ability of reproducing the
Gubser-Klebanov-Polyakov (GKP) prediction
\be
E\sim 2\,\sqrt{n}\,\lambda^{1/4},
\ee
for the energy of level $n$ massive string states as
$\lambda\to\infty$~\cite{Gubser:1998bc}.  The SBA equations are known
to generically agree with the GKP law, at least under mild reasonable
assumptions on the asymptotic behavior of Bethe momenta as
$\lambda\to\infty$ at finite $J$~\cite{Arutyunov:2004vx}.
Nevertheless, the results are reliable for large $J$ only which is the
limit where the SBA equations have been derived.

The easiest cases where the GKP law can be explicitly investigated
(determining also the dual level $n$) are the highest states in the
compact rank-1 subsectors $\mathfrak{su}(2)$ and
$\mathfrak{su}(1|1)$~\cite{Zarembo:2005ur,Arutyunov:2006av,Beccaria:2006td}.
For notational purposes we shall call these states {\em
antiferromagnetic} (AF) borrowing the wording from the
$\mathfrak{su}(2)$ case.  In a recent paper~\cite{Beccaria:2006td},
two of us proved that the SBA equations predict the following result
\be
\label{eq:old}
\frac{\Delta_{\mathfrak{su}(2)}^{\rm SBA}(L, \lambda)}{2L} =  \displaystyle\frac{1}{2}\,\lambda^{1/4} + {\cal O}(\lambda^0), \qquad
\frac{\Delta_{\mathfrak{su}(1|1)}^{\rm SBA}(L, \lambda)}{L} = \displaystyle \frac{1}{\sqrt{2}}\left(1-\frac{1}{L^2}\right)^{1/2}\, \lambda^{1/4} + {\cal O}(\lambda^0),
\ee
where $L$ denotes the number of Bethe momenta in both
sectors. Unfortunately, it is not easy to compare these results with
string theory calculations since the state dual to the AF operators is
not known. The only exception is the proposal~\cite{Roiban:2006jt} in
the $\mathfrak{su}(2)$ sector which, however, does not agree with the
GPK law.

The result Eq.~(\ref{eq:old}) is obtained at fixed $L$ and studying
the suitable solution of the SBA equations as $\lambda\to\infty$.
Since the SBA equations are valid for large $L$, one would like to
know how many terms in the $1/L$ expansion of Eq.~(\ref{eq:old}) are
correct.  Indeed, the correct procedure would require to take the
large $L$ limit of the SBA equations obtaining the functions
$\Delta^{AF}_{0,1}(\lambda)$ appearing in the expansion
\be
\frac{\Delta^{\rm SBA}(L, \lambda)}{L} =  \Delta^{AF}_0(\lambda) + \frac{1}{L}\,\Delta^{AF}_1(\lambda) + \dots.
\ee
Then, one could safely take the large $\lambda$ limit of each
term. This is what we denote the $L, \lambda~\to~\infty$ limit.
Unfortunately, it is not known how to solve the integral equations for
the Bethe roots distribution at large $L$ in neither sector. Also,
their strong coupling expansion is ambiguous and only the $\lambda,
L\to\infty$ limit, i.e. expanding in power of $\lambda^{-1/4}$ with
fixed $L$ and eventually expanding in 1/L, is currently
calculable. 

In principle, the SBA equations are only one possible discretization
of the classical string Bethe equations. Also, different gauge--fixed
formulation can lead to equivalent equations, although with their
special technical features.  A remarkable example is indeed described
in~\cite{Frolov:2006cc} where quantum SBA equations are derived
starting from the string action in the so-called uniform light-cone
gauge. This is the generalization of the usual flat space light-cone
gauge to the $AdS_5\times S^5$
case~\cite{Goddard:1973qh,Metsaev:2000yu,Arutyunov:2006ak,Kruczenski:2004kw,Kruczenski:2004cn,Arutyunov:2004yx,Arutyunov:2005hd,Frolov:2006cc,Itoyama:2006cg}.
Again, the equations are obtained starting from the leading
thermodynamical term in a (suitable light-cone) $1/J$ expansion and
discretising. In~\cite{Frolov:2006cc} the equations are matched to the
near-BMN corrections to pp-wave states fixing the leading dressing
phase.  Remarkably, a compact set of equations is obtained where the
dressing phase is somewhat reabsorbed.

The light-cone Bethe Ansatz equations (LCBA) recast the spectral
problem in an intriguing way and deserve in our opinion further
investigation.  In this paper, we analyze them working on the AF
states at large $\lambda$. We indeed show that the calculation
in~\cite{Beccaria:2006td} can be repeated in the LCBA framework
achieving much more insight. In particular, in the
$\mathfrak{su}(1|1)$ sector, we are able to solve them in the safe $L,
\lambda\to\infty$ limit clarifying the accuracy of our previous
calculation Eq.~(\ref{eq:old}).

\section{The light-cone Bethe Ansatz}
\label{sec:LC}

We briefly review the LCBA equations derived in~\cite{Frolov:2006cc}
to setup the notation. The uniform light-cone gauge is based on the
introduction of light-cone variables
\be
X_\pm = \frac{1}{2}(\varphi\pm t),
\ee
where $\varphi$ is an angle on $S^5$ conjugate to the angular momentum
$J$ and $t$ is the global time on $AdS_5$ conjugate to the energy
$E$. The gauge is fixed by the choice
\be
X_+ = \tau,\qquad p_+ = P_+ = \mbox{const},
\ee
where $p_+$ is conjugate to $x_-$. The world-sheet light-cone Hamiltonian is 
\be
H_{\rm lc} = -P_-,
\ee
and is a function of $P_+$. Expanding at large $P_+$ with
$\lambda/P_+^2$ fixed one recovers the BMN and near-BMN limit suitable
to study the pp-wave states. The two equations
\ba
E-J &=& H_{\rm lc}(P_+), \\
E+J &=& P_+,
\ea
lead to the following equation determining $E$
\be
E = J + H_{\rm lc}(E+J).
\ee
The results for pp-wave states in all rank-1 sectors  (including the non compact
$\mathfrak{sl}(2)$ case) are consistent at ${\cal O}(1/P_+)$ with the discrete equations
\be
\exp\left(i\,p_k\,\frac{P_++\mathfrak{s}\,M}{2}\right) = \mathop{\prod_{j=1}^M}_{j\neq k}\left(
\frac{x_k^+-x_j^-}{x_k^--x_j^+}\right)^\mathfrak{s},
\ee
where $\mathfrak{s} = -1, 0, 1$ in $\mathfrak{sl}(2)$, $\mathfrak{su}(1|1)$, and $\mathfrak{su}(2)$. The variables $x^\pm$ are
\be
x^\pm(p) = \frac{1}{4}\left(\cot\frac{p}{2}\pm i\right)\left(1+H_{\rm lc}(p)\right),
\ee
where
\be
H_{\rm lc}(p) = \sqrt{1+\frac{\lambda}{\pi^2}\sin^2\frac{p}{2}}.
\ee
In the next Sections we shall analyze in details the properties of the
highest energy solution of the above equation in the two cases
$\mathfrak{s} = 0, 1$. Of course, the energy $E$ must be identified
with the anomalous dimension $\Delta$ of the dual gauge invariant
operators.

\section{$\mathfrak{su}(1|1)$ sector}
\label{sec:su11}

\subsection{General features of the LCBA equations}

The light cone Bethe equations are particularly simple in the $\mathfrak{su}(1|1)$ sector and read
\be
\exp\left(i\,\frac{P_+}{2}\,p_k\right) = 1,
\ee
\be
P_- = \sum_{k=1}^M\sqrt{1+\frac{\lambda}{\pi^2}\sin^2\frac{p_k}{2}}.
\ee
where $P_\pm = \Delta\pm J$.  
To study the highest state and match the notation in \cite{Beccaria:2006td} we consider
\be
J = \frac{L}{2},\quad M=L,\quad L\in 2\mathbb{N}+1.
\ee
The equation for the Bethe momenta can be solved immediately and gives
\be
\label{eq:solved}
p_k = \frac{4\pi}{\Delta + L/2}\,n_k,\qquad n_k\in \zz.
\ee
The remaining equation determines $\Delta_L(\lambda)$
\be
\label{eq:delta11}
\Delta_L(\lambda) = \frac{L}{2} + \sum_{k=1}^L\sqrt{1+\frac{\lambda}{\pi^2}\sin^2\frac{2\pi\,n_k}{\Delta_L(\lambda)+L/2}}.
\ee
We solve this equation with $\{n_k\}$ in the symmetric range
\be
\{n_k\} = \left\{-\frac{L-1}{2}, \dots, 0, \dots, \frac{L-1}{2}\right\},
\ee
which uniquely selects the highest state~\cite{Arutyunov:2006av,Beccaria:2006td}. In App.~(\ref{app:unique})
we prove that the above equation admits a  unique solution $\Delta_L(\lambda)$ at fixed $L$.

\bigskip
The LCBA being derived at strong coupling, they should not be trusted
to yield a correct solution for $\Delta(\lambda)$ at weak coupling;
however, in the same spirit of~\cite{Arutyunov:2006av}, we present in
App.~(\ref{app:weak}) some results for its weak coupling expansion
which could be useful to compare with those from future
improved LCBA equations with a refined dressing.

\subsection{Strong coupling expansion in the $\lambda\to\infty$ limit at fixed $L$}

As explained in the Introduction, we begin our analysis of the LCBA
equations by studying the $\lambda, L\to \infty$ limit.  In other
words, we fix $L$ and take the large $\lambda$ limit, eventually
expanding in $1/L$.

Due to the simplicity of the equations, we can prove analyticity at
strong coupling, {\em i.e.} exclude non-analytic corrections to the
above relation as well as prove its convergence in a suitable
neighborhood of $\lambda = +\infty$. This is non trivial and indeed is
false in the $L\to\infty$ limit as we shall discuss later. The proof
of analyticity is reported in App.~(\ref{app:analytic}).

We can now systematically evaluate the perturbative strong coupling
coefficients.  We denote the large $\lambda$ expansion coefficients as
\be
\frac{\Delta_L(\lambda)}{L} = c_L\,\lambda^{1/4} + d_L  + e_L\,\lambda^{-1/4} + \cdots.
\ee
We easily find the leading term (from the theorem in
App.~(\ref{app:analytic}), see also Eq.(8.14) of \cite{Frolov:2006cc})
\be
c_L = \frac{2}{L}(\sum_{n_k>0} n_k)^{1/2} = \frac{2}{L} \sqrt{\frac{1}{2}\,\frac{L-1}{2}\,\frac{L+1}{2}} = \frac{1}{\sqrt 2}\left(1-\frac{1}{L^2}\right)^{1/2}.
\ee
in perfect agreement with \cite{Beccaria:2006td}. 

\bigskip
The NLO is also easy. The expansion of momenta is 
\be
p_k = \frac{\alpha_k}{\lambda^{1/4}}+\frac{\beta_k}{\lambda^{1/2}} + \dots
\ee
where
\be
\alpha_k = \frac{4\pi n_k}{L\,c_L},\qquad\beta_k = -\frac{4\pi n_k}{L\,c_L^2}\left(d_L + \frac{1}{2}\right).
\ee
On the other hand when $p\neq 0$ we have 
\be
\sqrt{1+\frac{\lambda}{\pi^2}\sin^2\frac{p_k}{2}} = \lambda^{1/4}\frac{1}{2\pi}|\alpha_k| + \frac{1}{2\pi}\beta_k\,\mbox{sign}\,\alpha_k.
\ee
Taking into account the term with $p=0$ we obtain
\be
\frac{\Delta_L(\lambda)}{L} = \lambda^{1/4}\frac{1}{2\pi\,L}\sum_k|\alpha_k| + \frac{1}{2\pi\,L}\sum_k\beta_k\,\mbox{sign}\,\alpha_k+\frac{1}{2}+\frac{1}{L}+\dots
\ee
Consistency requires as before
\be
\sum_k |n_k| = \frac{L^2}{2} c_L^2,
\ee
but also
\ba
d_L &=& \frac{1}{2}+\frac{1}{L} + \frac{1}{2\pi\,L}\sum_k \frac{-4\pi n_k}{L\,c_L^2}\left(d_L + \frac{1}{2}\right)\mbox{sign}\, n_k = \\
&=& \frac{1}{2}+\frac{1}{L} -\frac{2}{L^2\,c_L^2} \sum_k |n_k|\left(d_L + \frac{1}{2}\right) = \frac{1}{L}-d_L. \nonumber
\ea
Hence,
\be
d_L = \frac{1}{2L}.
\ee

\bigskip
The NNLO is more involved. After some calculations it  reads
\ba
e_L &=& \frac{1}{4\sqrt{2}}\left(1+\frac{1}{L}\right)^2 \left(1-\frac{1}{L^2}\right)^{-1/2}
-\frac{\pi^2}{12\sqrt{2}}\left(1-\frac{1}{L^2}\right)^{1/2} + \\
&&
+ \frac{1}{4\sqrt{2}} \left(1-\frac{1}{L^2}\right)^{1/2}\,\sum_{k=1}^{\frac{L-1}{2}}\frac{1}{k}. \nonumber
\ea
At large $L$, $e_L \sim \ln L/(4\sqrt 2)$ and does not admit a finite
limit as $L\to\infty$. In the next Section we shall discuss this
important point. It is worthwhile to emphasize that the NLO
contributions to the dressing factor can in principle modify this
term.  Hence, it could be correct if the LCBA equations turned out to
reabsorb the full dressing phase or, what is more natural, it would
not be reliable if the LCBA equations required additional corrections.

\bigskip
As a consistency check of the calculation, the above expansion is confirmed by the numerical solution of the equation for $\Delta$ as illustrated in Fig.~(\ref{fig:exp11})
where we show the constant values approached at large $\lambda$ by the difference
\be
\lambda^{1/2}\,\left(\frac{\Delta_L(\lambda)}{L} - c_L\,\lambda^{1/4} - d_L  - e_L\,\lambda^{-1/4}\right),
\ee
at various $L$.

\bigskip
To summarize, the main result of this Section is the expansion 
\be
\label{eq:summary1}
\lambda \to\infty,\qquad\qquad \frac{\Delta_L(\lambda)}{L} = \frac{1}{\sqrt{2}}\left(1-\frac{1}{L^2}\right)^{1/2}\,\lambda^{1/4} + \frac{1}{2L} + e_L\,\lambda^{-1/4} + \dots.
\ee
If one expands in $1/L$, then only the ${\cal O}(L^0)$ and ${\cal
O}(L^{-1})$ terms are reliable because the LCBA equations are derived
at first order in $1/P_+$.  However, $e_L$ has not a finite limit as
$L\to\infty$ as a hint of the fact that the physically meaningful
limit is the opposite one $L, \lambda\to\infty$.  In the next Section,
we shall confirm the calculation leading to (the first two terms of )
Eq.~(\ref{eq:summary1}) by an independent calculation using the SBA
equations in the same limit. Later, we shall discuss the opposite $L,
\lambda\to\infty$ case comparing it with Eq.~(\ref{eq:summary1}).

\section{Improved calculation in the SBA framework}
\label{sec:dressing}

As an independent check of Eq.~(\ref{eq:summary1}), we can repeat the calculation within the SBA equations. We briefly recall 
some information about the strong coupling expansion of the dressing factor. Then, we show that the leading term is enough to 
reproduce the first two terms in Eq.~(\ref{eq:summary1}). Finally we do the computation, finding full agreement.

\subsection{The dressing factor at strong coupling}

The quantum string Bethe Ansatz equations can be written~\cite{Arutyunov:2004vx}
\be
e^{i\,p_i\,L} = \prod_{j\neq i}^M S_{ij},\qquad
S_{ij} = \left(\frac{x_i^+-x_j^-}{x_i^--x_j^+}\right)^\mathfrak{s}\,
\frac{\displaystyle 1-\frac{\lambda}{16\pi^2}\frac{1}{x_i^+\,x_j^-}}
{\displaystyle 1-\frac{\lambda}{16\pi^2}\frac{1}{x_i^-\,x_j^+}}\,e^{i\,\vartheta_{ij}},
\ee
where $\vartheta$ is the universal dressing factor.
In the various rank-1 sectors we have
\be
\mathfrak{s} = \begin{array}{ccc}
1 & 0 & -1\\
\{\mathfrak{su}(2) & \mathfrak{su}(1|1) & \mathfrak{sl}(2)\}
\end{array},
\qquad L = J + \frac{\mathfrak{s}+1}{2}\,M.
\ee
The variables $x^\pm$ are again
\be
x^\pm = \frac{e^{\pm i\, p/2}}{4\,\sin(p/2)}\left(1+\sqrt{1+\frac{\lambda}{\pi^2}\sin^2\frac{p}{2}}\right).
\ee
We now introduce the variables
\be
\zeta = \frac{2\pi}{\sqrt\lambda},\qquad 
\widetilde x^\pm = 2\,\zeta\,x^\pm.
\ee
The scattering phase can be written~\cite{Arutyunov:2006iu}
\be
\vartheta_{ij} = \frac{1}{\zeta}\sum_{r\ge 2}\sum_{n\ge 0} c_{r, r+1+2n}(\zeta)\,(q_r(\widetilde x_i)\,q_{r+1+2n}(\widetilde x_j) - (i\leftrightarrow j)),
\ee
where the local charges are 
\be
q_r(\widetilde x) = \frac{i}{r-1}\left(\frac{1}{(\widetilde{x}^+)^{r-1}}-\frac{1}{(\widetilde{x}^-)^{r-1}}\right).
\ee
The first two terms in the $\zeta$-expansion of $c_{r,s}$ are
\be
c_{r,s} = \delta_{r+1, s}-\zeta\,\frac{4}{\pi}\frac{(r-1)(s-1)}{(r+s-2)(s-r)} + {\cal O}(\zeta^2).
\ee
The scattering phase can be organized as 
\ba
\vartheta_{ij} &=& \frac{1}{\zeta}[\chi_{ij}^{--}-\chi_{ij}^{-+}-\chi_{ij}^{+-}+\chi_{ij}^{++}-(i\leftrightarrow j)], \\
\chi_{ij}^{\sigma\sigma'} &=& \chi\left(\frac{4\pi}{\sqrt\lambda}\, x_i^\sigma, \frac{4\pi}{\sqrt\lambda}\, x_j^{\sigma'}\right).
\ea
The function $\chi$ is expanded at strong coupling as:  
\be
\label{eq:chi}
\chi = \sum_{n\ge 0} \chi_n\,\zeta^n,
\ee
and the first two terms $\chi_{0,1}$ can be given in closed form~\cite{Arutyunov:2006iu}
\be
\chi_0(x,y) = -\frac{1}{y}-\frac{xy-1}{y}\log\frac{xy-1}{xy},
\ee
\ba
\chi_1(x,y) &=& \frac{1}{\pi}\left[\log\frac{y-1}{y+1}\log\frac{x-1/y}{x-y} + \right.\nonumber \\
&&  + \mbox{Li}_2\left(\frac{\sqrt{y}-1/\sqrt{y}}{\sqrt{y}-\sqrt{x}}\right)-\mbox{Li}_2\left(\frac{\sqrt{y}+1/\sqrt{y}}{\sqrt{y}-\sqrt{x}}\right) \\
&& \left. + \mbox{Li}_2\left(\frac{\sqrt{y}-1/\sqrt{y}}{\sqrt{y}+\sqrt{x}}\right)-\mbox{Li}_2\left(\frac{\sqrt{y}+1/\sqrt{y}}{\sqrt{y}+\sqrt{x}}\right)\right]\nonumber
\ea
Notice we cannot read trivially the powers of $\zeta$ from Eq.~(\ref{eq:chi}) because $\lambda$ appears
non trivially in the arguments of $\chi_n$ as well as in the expression of $x^\pm$.

\subsection{Subleading corrections to the SBA equations}

Let us expand at large $\lambda$ the Bethe momenta 
\be
p_k = \frac{\alpha_k}{\lambda^{1/4}} + \frac{\alpha'_k}{\lambda^{1/2}} + \cdots.
\ee
The dressing phase at LO and NLO can also be expanded (for $\alpha_k>0$) with the result
\ba
\vartheta^{\rm LO}_{kj} &=& \frac{\alpha_k\alpha_j}{2\pi} + \frac{1}{\lambda^{1/4}}\left[
\alpha_k-\frac{2}{3}\alpha_j+\frac{\alpha'_k\alpha_j}{3\pi}+\frac{|\alpha_j|}{3}-\frac{\alpha'_k |\alpha_j|}{6\pi} + \frac{\alpha_k \alpha'_j}{2\pi}\right], \\
\vartheta^{\rm NLO}_{kj} &=& {\cal O}(\lambda^{-1/2}).
\ea
This means that the coefficients $\alpha'$ can be determined from the LO only, {\em i.e.} without involving $\chi_1$ and its
complicated analytic structure.

Let us now compute the subleading corrections in the SBA equations that explicitly involve the LO dressing factor.
The SBA equations in logarithmic form are 
\be
p_k\,L = 2\pi n_k + \sum_{j\neq k} \left(\vartheta_{jk} -i\,\log\frac{
\displaystyle 1- \frac{\lambda}{16\pi^2}\frac{1}{x^+_j x^-_k}}
{\displaystyle 1- \frac{\lambda}{16\pi^2}\frac{1}{x^-_j x^+_k} }\right)
\ee
With the previous notation for the sums and exploiting antisymmetry of the term in brackets we find 
\be
\sum_{k\in P} p_k\,L = 2\pi \sum_{k\in P} n_k + \sum_{k\in P}\sum_{j\in M} \left(\vartheta_{jk} -i\,\log\frac{
\displaystyle 1- \frac{\lambda}{16\pi^2}\frac{1}{x^+_j x^-_k}}
{\displaystyle 1- \frac{\lambda}{16\pi^2}\frac{1}{x^-_j x^+_k} }\right)
\ee
For $k\in P$ and $j\in M$ we have at large $\lambda$
\be
-i\,\log\frac{
\displaystyle 1- \frac{\lambda}{16\pi^2}\frac{1}{x^+_j x^-_k}}
{\displaystyle 1- \frac{\lambda}{16\pi^2}\frac{1}{x^-_j x^+_k} } = \frac{1}{2}(\alpha_j-\alpha_k)\,\frac{1}{\lambda^{1/4}} + \cdots
\ee
At leading order we insert the first term of the expansion of $\vartheta^{\rm LO}$ and obtain 
\be
0 = 2\pi\sum_{k\in P} n_k +  \frac{1}{2\pi}\sum_{k\in P}\sum_{j\in M} \alpha_k\,\alpha_j.
\ee
Using parity invariance we recover the known result
\be
S\equiv \sum_{k\in P}\alpha_k = \frac{\pi}{\sqrt{2}}\,(L^2-1)^{1/2}.
\ee
The next correction is determined from the ${\cal O}(\lambda^{-1/4})$ terms. The equation is 
\be
L\,\sum_{k\in P}\alpha_k = \sum_{k\in P}\sum_{j\in M}\left\{
\frac{1}{2}(\alpha_j-\alpha_k)+\alpha_k-\frac{2}{3}\alpha_j+\frac{\alpha'_k\alpha_j}{3\pi}+\frac{|\alpha_j|}{3}-\frac{\alpha'_k |\alpha_j|}{6\pi} + \frac{\alpha_k \alpha'_j}{2\pi}
\right\}
\ee
Evaluating the sums and defining also 
\be
S' = \sum_{k\in P}\alpha_k',
\ee
we easily obtain (again exploiting parity invariance of the Bethe momenta)
\be
L\,S = \frac{L-1}{2}\,S-\frac{1}{\pi} S\,S'\qquad\la\qquad S' = -\pi\frac{L+1}{2}.
\ee
The asymptotic expansion of the anomalous dimension is 
\ba
\Delta_L(\lambda) &=& \frac{L}{2} + 1 + 2\sum_{k\in P}\sqrt{1 + \frac\lambda{\pi^2}\sin^2\frac{p_k}{2}} \\
&\la& \frac{\lambda^{1/4}}{\pi}\,S + \frac{1}{\pi}\,S' + \frac{L+2}{2} + {\cal O}(\lambda^{-1/4}).
\ea
Replacing the values of $S$ and $S'$ we obtain 
\be
\frac{\Delta_L(\lambda)}{L} = \frac{1}{\sqrt{2}}\left(1-\frac{1}{L^2}\right)^{1/2}\,\lambda^{1/4} + \frac{1}{2L} +  {\cal O}(\lambda^{-1/4}),
\ee
in full agreement with the first two terms in Eq.~(\ref{eq:summary1}).

\section{The $\mathfrak{su}(1|1)$ sector at $L\to\infty$ limit at fixed $\lambda$}
\label{sec:Linf}

In the previous Sections we have established Eq.~(\ref{eq:summary1})
which is the $\lambda, L\to \infty$ strong coupling expansion of the
anomalous dimension. In this Section, we discuss the $1/L$ expansion
of the equation determining $\Delta$ as well as its (correct) $L,
\lambda\to\infty$ limit. The main point is that in the
$\mathfrak{su}(1|1)$ sector the LCBA equations for the Bethe momenta
are immediately solved by Eq.~(\ref{eq:solved}). Thus, we do not need
any integral equation for the Bethe root distribution and we simply
have to take the $1/L$ expansion of a trascendental equation for
$\Delta$ itself.

\bigskip
We start with the LCBA equation that we write as 
\be
\Delta_L(\lambda) = \frac{L}{2} + 1 + 2\,\sum_{k=1}^{\frac{L-1}{2}} \sqrt{1+\frac{\lambda}{\pi^2}\sin^2\frac{2\pi k}{\Delta_L(\lambda)+\frac{L}{2}}}.
\ee
The sum in the r.h.s. can be evaluated by  applying the Euler-MacLaurin summation formula ($B_k$ are Bernoulli numbers) 
\be
\label{eq:euler}
\sum_{k=1}^{N-1} f(k) = \int_0^N f(x) dx -\frac{1}{2}\left[f(0)+f(N)\right]  
+ \sum_{k\ge 1}\frac{B_{2k}}{(2k)!}\left[f^{(2k-1)}(N)-f^{(2k-1)}(0)\right],
\ee
where we notice that in our case $f^{(2k-1)}(0) = 0$. The Euler-MacLaurin formula provides an asymptotic 
expansion in powers of $1/L$ because the $k$-th term in the last contribution to Eq.~(\ref{eq:euler}) scales like $1/L^{2k-1}$.
Our strategy will be that of solving the LCBA equation order by order in $1/L$ in this asymptotic expansion.
As discussed in App.~(\ref{app:plana}) the expansion is expected to be only asymptotic in the Poincar\'e sense.
This is not surpring since the $1/L$ expansion computes the various loop corrections in the $\sigma$-model and zero radius of 
convergence is a common feature in perturbation theory of non-trivial field theories.
\bigskip
Writing 
\be
\Delta_L(\lambda) = L\left(u-\frac{1}{2}\right) + z_1 + \frac{1}{L}\, z_2 + \cdots
\ee
and expanding, we obtain the leading order result
\be
\label{eq:lo}
u = 1 + \frac{u}{\pi}\,\mbox{E}\left(\frac{\pi}{u}, -\frac{\lambda}{\pi^2}\right),
\ee
where $\mbox{E}(z, m)$ is the standard incomplete elliptic integral of the second kind
\be
\mbox{E}(z, m) = \int_0^z\sqrt{1-m\,\sin^2\theta}\,d\theta.
\ee
The NLO and NNLO corrections are determined by $u$ and are given by
\ba
z_1 &=& 0, \\
z_2 &=& -\frac{\lambda}{12\,\pi}\,\sin\frac{2\pi}{u}\frac{1}{\sqrt{1+\frac{\lambda}{\pi^2}\sin^2\frac{\pi}{u}}
(1+\sqrt{1+\frac{\lambda}{\pi^2}\sin^2\frac{\pi}{u}})}.
\ea
Remarkably, the ${\cal O}(1/L)$ correction vanishes. We can now expand $u(\lambda)$ at large $\lambda$. As a consistency check, 
we also report in App.~(\ref{app:Linfweak}) the expansion at small $\lambda$.

\bigskip
Setting
\be
x = \frac{\pi}{u},\qquad t = \frac{\lambda}{\pi^2},
\ee
we have to solve for $t\to \infty$ the equation
\be
\pi = x + \mbox{E}(x, -t).
\ee
The expansion of this equation at $t\to +\infty$ and $x = {\cal O}(t^{-1/4})$ is not at all trivial.
It is worked out in App.~(\ref{app:expansion})  with the result
\be
 \mbox{E}(x, -t) = (1-\cos\,x)\sqrt{t} + \frac{1}{4\sqrt{t}}\left(1+\log(16 t) + 2\log\tan\frac{x}{2}\right) + \dots
\ee
Using this expansion, the solution of the above equation turns out to be
\be
x(t) = \sqrt{2\pi} t^{-1/4} + \frac{1}{24\sqrt{2\pi}}\,t^{-3/4}(-3\log t -6\log(8\pi)+6+4\pi^2) + \dots
\ee
Replacing in 
\be
\lim_{L\to\infty}\frac{{\Delta_L(\lambda)}}{L} = \frac{\pi}{x}-\frac{1}{2},
\ee
we obtain 
\ba
\lim_{L\to\infty}\frac{{\Delta_L(\lambda)}}{L} &=& u-\frac{1}{2} = \frac{1}{\sqrt{2}}\,\lambda^{1/4} + \\
&+& \frac{1}{48\sqrt{2}}\,\lambda^{-1/4}\,(3\log\lambda + 18\log 2 + 18-4\pi^2) + \nonumber \\
&+& \phantom{\frac{1}{2}} {\cal O}(\lambda^{-1/2}\,\log\lambda). \nonumber 
\ea
Replacing the strong coupling expansion of $u$ we obtain 
\be
z_2 = -\frac{1}{6\sqrt{2}}\,\lambda^{1/4} + {\cal O}(\lambda^{-1/4}\,\log\lambda).
\ee
In summary, we have found in the $L,\lambda\to \infty$ limit:
\ba
\label{eq:ok}
\lefteqn{
\qquad \frac{{\Delta_L(\lambda)}}{L} = \frac{1}{\sqrt{2}}\,\lambda^{1/4} +} &&  \\
&+& \frac{1}{48\sqrt{2}}\,\lambda^{-1/4}\,(3\log\lambda + 18\log 2 + 18-4\pi^2) + {\cal O}(\lambda^{-1/2}\,\log\lambda) + \nonumber \\
&+& \frac{1}{L^2}\left(-\frac{1}{6\sqrt{2}}\,\lambda^{1/4} + {\cal O}(\lambda^{-1/4}\,\log\lambda)\right) + \dots.\nonumber
\ea
A comparison with Eq.~(\ref{eq:summary1}) shows that the two limits in $\lambda$ and $L$
do not commute. The equations are valid in the order $L, \lambda\to\infty$ where the result is Eq.~(\ref{eq:ok}).
{\em Only the leading term both in $L$ and $\lambda$}  
is independent on the order. Similar results in the Hubbard model formulation of the gauge BA are illustrated in~\cite{Feverati:2006hh}.

Actually, Eq.~(\ref{eq:ok}) contains an additional information beyond this term. The second line is 
in principle affected by the NLO strong coupling terms in the dressing factor that,
honestly,  is not expected to be taken into account in the LCBA equations. Also, the $1/L^2$ term in the third line
is beyond the validity of the equations that are fixed by looking at ${\cal O}(1/P_+)$ corrections.  
Nevertheless, we have proved that $z_1=0$. Thus, our result reads
\be
L, \lambda\to\infty,\qquad \frac{{\Delta_L(\lambda)}}{L} = \frac{1}{\sqrt{2}}\,\lambda^{1/4} + {\cal O}\left(\frac{\lambda^{-1/4}}{L}\right)  + {\cal O}\left(\frac{\lambda^{1/4}}{L^2}\right) .
\ee
Within this precision, the discrepancy with the $\lambda, L\to\infty$ limit is localized in the $1/(2L)$ term appearing in Eq.~(\ref{eq:summary1}).

\section{The $\mathfrak{su}(2)$ sector}
\label{sec:su2}

\subsection{General features of the LCBA equations}

The LCBA equations read in this sector
\be
\exp\left(i\,p_k\,\frac{P_++M}{2}\right) = \prod_{j\neq k} \frac{x^+_k-x^-_j}{x^-_k-x^+_j}
\ee
where
\be
x^\pm = \frac{1}{4}\left(\cot\frac{p}{2}\pm i\right)\left(1+\sqrt{1+\frac{\lambda}{\pi^2}\sin^2\frac{p}{2}}\right),
\ee
where again
\ba
P_+ &=& \Delta + J, \\
P_- &=& \Delta - J = \sum_{k=1}^M \sqrt{1+\frac{\lambda}{\pi^2}\sin^2\frac{p_k}{2}}.
\ea
We are interested in the sector of operators with $2L$ fields and zero angular momentum. So $M=L$, $J=2L-M = L$.

\subsubsection{$\lambda=0$}

Let us first discuss the case $\lambda=0$. In this limit
\be
\frac{1}{2}(P_++M) = J+M = 2L.
\ee
We define
\ba
u &=& \frac{1}{2} \cot\frac{p}{2}, \\
p &=& 2 \arctan\frac{1}{2u}.
\ea
The map is 1-1 with $u\in\rr$ and $p\in(-\pi, \pi)$. By standard manipulations we arrive at 
\be
2\,\sum_{j\neq k} \arctan(u_k-u_j) -4\,L\,\arctan(2 u_k) = 2\pi\,J_k,
\ee
where the correct choice of Bethe quantum numbers for the AF state is 
\be
J_k = \{-\frac{L-1}{2}, -\frac{L-3}{2}, \dots, \frac{L-3}{2}, \frac{L-1}{2}\}. 
\ee
The associated solution has the first $L/2$ $u>0$ and the other negative. 
This Bethe equation can be recast in terms of the  $p$ variables and reads
\be
2\,\sum_{j\neq k} \arctan\left(\frac{1}{2}\cot\frac{p_k}{2}-\frac{1}{2}\cot\frac{p_j}{2}\right) + 2\,L\,p_k = 2\pi\,R_k
\ee
with 
\be
R_k = \{\frac{L+1}{2}, \frac{L+3}{2}, \dots, L-\frac{1}{2}\}\cup (\mbox{the opposite list}).
\ee

\subsubsection{$\lambda \neq 0$}

We now take $\lambda>0$. We simply have to replace
\be
2L\la\,\frac{{\Delta_L(\lambda)}+2L}{2}
\ee
and use the more complicated form of $x^\pm$. The result is 
\be
2\,\sum_{j\neq k} \arctan\left(X_{kj}\right) + \frac{{\Delta_L(\lambda)}+2L}{2}\,p_k = 2\pi\,R_k, 
\ee
\be
{\Delta_L(\lambda)} = L + \sum_{k=1}^L \sqrt{1+\frac{\lambda}{\pi^2}\sin^2\frac{p_k}{2}}
\ee
with the above $\{R_k\}$ and where
\be
X_{kj} = \frac{\cot\frac{p_k}{2}(1+\sqrt{1+\frac{\lambda}{\pi^2}\sin^2\frac{p_k}{2}})-\cot\frac{p_j}{2}(1+\sqrt{1+\frac{\lambda}{\pi^2}\sin^2\frac{p_j}{2}})}
{2+\sqrt{1+\frac{\lambda}{\pi^2}\sin^2\frac{p_k}{2}}+\sqrt{1+\frac{\lambda}{\pi^2}\sin^2\frac{p_j}{2}}}.
\ee

\subsection{The $\lambda\to\infty$ limit at fixed $L$}

The numerical solution of the LCBA equations at fixed $L$ by means of the Newton algorithm~\cite{Newton} is perfectly feasible as discussed in full details in \cite{Beccaria:2006td}. 
In Fig.~(\ref{fig:su2sol}), we show the case $L=10$ and the scaled momenta $\lambda^{1/4}\,p_k$.
In the same Figure, we have also shown the analytical prediction for the asymptotic $p$ as derived below.

We can assume $p_k\sim \alpha_k\,\lambda^{-1/4}$. If $\alpha_i>0$ and $\alpha_j<0$ we have at large $\lambda$
\be
X_{ij} \to \frac{4\lambda^{1/4}}{\alpha_i-\alpha_j}\to +\infty
\ee
If instead  $\alpha_i, \alpha_j >0$ we find
\be
X_{ij} \to -\frac{4\pi}{\alpha_i\alpha_j}\frac{\alpha_i-\alpha_j}{\alpha_i+\alpha_j}.
\ee
The Bethe equations reduce to 
\be
\frac{\pi\,L}{2}-2\sum_{\alpha_j>0,\ j\neq i}\arctan\left(\frac{4\pi}{\alpha_i\alpha_j}\frac{\alpha_i-\alpha_j}{\alpha_i+\alpha_j}\right)
+\frac{1}{2\pi}\alpha_i\sum_{\alpha_j>0}\alpha_j = 2\pi\,R_i.
\ee
This equation determines the $\alpha_i>0$. For instance, for $L=10$ we find
\ba
\alpha_1 &=& 3.51948258944006628701335069602, \nonumber \\
\alpha_2 &=& 4.99794847442355681996436161771,\nonumber \\
\alpha_3 &=& 6.34122560474632117268892612476, \\
\alpha_4 &=& 7.63958729635505788529227272867, \nonumber \\
\alpha_5 &=& 8.91768257093293021966752266564,\nonumber
\ea
which are the values appearing in the previous figure. The distribution of positive $\alpha_k$ for $L=150$ is 
shown in Fig.~(\ref{fig:alpha}).

\bigskip
The strong coupling expansion of ${\Delta_L(\lambda)}$ is thus again
\be
\frac{{\Delta_L(\lambda)}}{2 L} = c_L\,\lambda^{1/4}+d_L + {\cal O}(\lambda^{-1/4})
\ee
The analogous expansion for the Bethe momenta is
\be
p_k = \frac{\alpha_k}{\lambda^{1/4}} + \frac{\beta_k}{\lambda^{1/2}} + \cdots
\ee
and note that there is symmetry $p\to -p$ in the solution for the highest state.
Expanding, we find ($\epsilon_x = \mbox{sign}\, x$)
\ba
\frac{{\Delta_L(\lambda)}}{2L} &=& \frac{1}{2} + \frac{1}{2L} \sum_{k=1}^M\sqrt{1+\frac{\lambda}{\pi^2}\sin^2\frac{p}{2}} = \\
&=& \frac{\lambda^{1/4}}{4\pi L}\sum_k |\alpha_k| + \frac{1}{4\pi L}\sum_k \epsilon_{\alpha_k} \beta_k + \frac{1}{2} + \dots = \nonumber \\
&=& \frac{\lambda^{1/4}}{2\pi L}\sum_{k\in P} \alpha_k + \frac{1}{2\pi L}\sum_{k\in P} \beta_k + \frac{1}{2} + \dots, \nonumber
\ea
where $P$ is the set of $k$ such that $\alpha_k>0$.

\bigskip
The Bethe equation can be written in the $\lambda\to\infty$ limit and for $i\in P$
\be
2\,\sum_{j\in P} \arctan X_{ij} + \sum_{j\not\in P} \left(\pi-\frac{1}{2} (\alpha_i-\alpha_j) \lambda^{-1/4} + \dots\right)
+ \frac{\alpha_i}{4\pi}\sum_j |\alpha_j| + 
\ee
\be
 + \frac{1}{2}\left\{
\frac{\beta_i}{2\pi}\sum_j |\alpha_j| + 3\,L\,\alpha_i + \frac{\alpha_i}{2\pi}\sum_j \epsilon_{\alpha_j}\,\beta_j\right\}\lambda^{-1/4} = 2\pi R_i.
\ee
and using the parity symmetry
\be
2\,\sum_{j\in P} \arctan X_{ij} + \sum_{j\in P} \left(\pi-\frac{1}{2} (\alpha_i+\alpha_j) \lambda^{-1/4} + \dots\right)
+ \frac{\alpha_i}{2\pi}\sum_{j\in P} \alpha_j + 
\ee
$$
 + \frac{1}{2}\left\{
\frac{\beta_i}{\pi}\sum_{j\in P} \alpha_j + 3\,L\,\alpha_i + \frac{\alpha_i}{\pi}\sum_{j\in P} \beta_j\right\}\lambda^{-1/4} = 2\pi R_i.
$$
We now sum over $i\in P$ and due to
\be
\sum_{i,j\in P}\arctan X_{ij} = 0,
\ee
we obtain 
\be
\frac{\pi L^2}{4} - \frac{L}{2}\sum_{i\in P} \alpha_i\,\lambda^{-1/4}
+ \frac{1}{2\pi}(\sum_{i\in P} \alpha_i)^2
\ee
$$
 + \left\{
\frac{1}{\pi}\sum_{i\in P} \beta_i\, \sum_{j\in P} \alpha_j + \frac{3}{2}\,L\,\sum_{i\in P}\alpha_i \right\}\lambda^{-1/4} = 2\pi \sum_{R_i>0} R_i.
$$
Collecting terms, we find the leading order
\be
\frac{1}{2\pi}(\sum_{\alpha_i>0}\alpha_i)^2 = 2\pi\,\sum_{R_i>0}\, R_i-\frac{\pi L^2}{4} =
2\pi\,\frac{3L^2}{8}-\frac{\pi L^2}{4} = \frac{\pi \, L^2}{2}
\ee
Hence, 
\be
\sum_{\alpha_i>0}\alpha_i = \pi\,L,\qquad\la\qquad c_L = \frac{1}{2}.
\ee
Also, the NLO terms give
\be
\frac{1}{\pi}\sum_{i\in P}\beta_i = -L,\qquad\la\qquad d_L = 0.
\ee
In summary,
\be
\lambda\to\infty,\qquad \frac{{\Delta_L(\lambda)}}{2 L} = \frac{1}{2}\,\lambda^{1/4}+ 0 + {\cal O}(\lambda^{-1/4}).
\ee
This result can also be obtained in the SBA framework by repeating the calculation we did in 
the $\mathfrak{su}(1|1)$ sector. Unfortunately, here we are not able to find the strong coupling limit of the LCBA equations 
at large $L$, exactly as with the SBA equations. From our experience in the  $\mathfrak{su}(1|1)$ it seems very reasonable to claim 
that the leading term $1/2\,\lambda^{1/4}$ is independent on the order of limits.

\section{Conclusions}

The story of AdS/CFT duality is vexed by discrepancies related to the different limits in which 
calculations can be performed under control on the two sides of the correspondence. This is 
usually considered a weak coupling problem. In BMN limits, one takes a large $R$-charge $J$ and 
't Hooft coupling $\lambda$ to control the string side. When going to small $\lambda' = \lambda/J^2$
it is possible to compare with gauge theory perturbative calculations, but this is just one of the infinite 
directions along which $\lambda$ and $J$ can grow. 

In this paper, we have considered these problems from another perspective wondering whether 
the large $\lambda$ and $L$ region is free of ambiguities. We have shown that this is true only at leading order.
Actually, this is a problem which is not immediately related to the AdS/CFT correspondence. Instead, it seems to be a 
genuine feature of the string Bethe Ansatz equations which are derived not only assuming that both $\lambda$ and $L$ are large, 
but also taking the two limits in a precise order. We have shown that the anomalous dimensions of the highest states in the compact 
rank-1 sectors do depend on the order of limits beyond the leading term. This is not at all surprising, but seems to us 
an important warning.

Our results have been possible due to the particular simplicity of the light-cone string Bethe Ansatz equations 
in the fermionic $\mathfrak{su}(1|1)$ sector where the Bethe roots distribution is trivial for all $L$. 
Gauge independence of the anomalous dimensions suggests that the result should hold also for the standard equations 
with the AFS phase, although we could not prove this statement in that context.

In conclusion, we remark that although rather special, highest states
appears to be an interesting island in the moduli space of the $\ads$
superstring, complementary to pp-wave and spinning string states.
Indeed, our limited investigation has revealed some subtleties in the
structural properties of its quantum Bethe Ansatz equations
enlightening with explicit calculations the detailed way in which the
Gubser-Klebanov-Polyakov law is reproduced.

\acknowledgments
We would like to thank A. Tseytlin for drawing our attention to the LCBA. 

\newpage

\appendix

\section{Existence and unicity of ${\Delta_L(\lambda)}$ in the $\mathfrak{su}(1|1)$ sector}
\label{app:unique}

\begin{theorem}
The light cone equation Eq.~(\ref{eq:delta11}) for ${\Delta_L(\lambda)}$ at fixed $L$ admits a unique solution.
\end{theorem}
\proof{
We write the equation in the form 
\be
\label{eq:A1}
{\Delta_L(\lambda)} = \frac{L}{2}+1+2\sum_{k=1}^{\frac{L-1}{2}}\sqrt{1+\frac{\lambda}{\pi^2}\sin^2\frac{2\pi\,k}{{\Delta_L(\lambda)}+L/2}}
\ee
The equation implies
\be
{\Delta_L(\lambda)} \ge \frac{L}{2}+1+2\,\frac{L-1}{2} = \frac{3L}{2}.
\ee
The derivative of the square root is
\be
\frac{d}{d{\Delta_L(\lambda)}}\sqrt{1+\frac{\lambda}{\pi^2}\sin^2\frac{2\pi\,k}{{\Delta_L(\lambda)}+L/2}} = -\frac{k\,\lambda}{\pi}\,
\frac{\sin\frac{4\pi k}{{\Delta_L(\lambda)}+\frac{L}{2}}}{({\Delta_L(\lambda)}+\frac{L}{2})^2\sqrt{1+\frac{\lambda}{\pi^2}\sin^2\frac{2\pi\,k}{{\Delta_L(\lambda)}+L/2}}}
\ee
When ${\Delta_L(\lambda)}\ge 3L/2$ and $1\le k \le (L-1)/2$, the above
expression is negative. Hence the right hand side of Eq.~\ref{eq:A1} is
monotonically decreasing with ${\Delta_L(\lambda)}$. We conclude that
there is always a unique intersection with the left hand side.  }

\newpage
\section{Weak coupling expansion in the $\mathfrak{su}(1|1)$ sector}
\label{app:weak}

It is clear that the function ${\Delta_L(\lambda)}$ (at fixed $L$) is holomorphic in $\lambda$ in a 
neighborhood of $\lambda=0$ by the analytic implicit function theorem~\cite{Gunning}.
We then expand
\be
\frac{{\Delta_L(\lambda)}}{L} = \sum_{n=0}^\infty \gamma^{(n)}_L\,\left(\frac{\lambda}{\pi^2}\right)^n
\ee
With some effort, the various coefficients can be evaluated analytically. The first is trivial
\be
\gamma_L^{(0)} = \frac{3}{2}.
\ee
The next coefficient is 
\be
\gamma_L^{(1)} = \frac{1}{L}\sum_{n=0}^{\frac{L-1}{2}}\sin^2\frac{n\,\pi}{L} = \frac{1}{4},\qquad (L\in 2\mathbb{N}+1)
\ee
We have computed the analytical expression of the  next two coefficients and it reads
\ba
\gamma_L^{(2)} &=& -\frac{1}{64}\left(3+\frac{2\pi}{L\,\sin\frac{\pi}{L}}\right), \\
\gamma_L^{(3)} &=& \frac{1}{1024\,L^2}\,\frac{1}{\cos^2\frac{\pi}{2L}}\left[
20\,L^2\,\cos^2\frac{\pi}{2L} + \pi\,L\,\cot\frac{\pi}{2L}\left(9+\frac{1}{\cos\frac{\pi}{L}}\right) + 2\pi^2\right].
\ea
Their expansion at large $L$ is 
\ba
\label{zzz}
\gamma_L^{(2)} &=& -\frac{5}{64} -\frac{1}{192}\,\frac{\pi^2}{L^2} + \dots, \\
\gamma_L^{(3)} &=& \frac{5}{128} +\frac{19}{3072}\,\frac{\pi^2}{L^2} + \dots. \nonumber
\ea
The other coefficients $\{\gamma^{(n)}_L\}_{n\ge 4}$ are more and more involved functions of $L$. Their expression is not enlightening.

Starting from $\gamma_L^{(2)}$ the expansion coefficients depend on $L$. This is in sharp contrast with what is obtained in the 
usual conformal gauge. There, all coefficients $\gamma_L^{(n)}$ are $L$-independent for a suitably large (but finite) $L$~\cite{Arutyunov:2006av,Beccaria:2006td}. 
Beside and more remarkably, the disagreement with perturbative gauge theory starts at two loops. This is a simple fact that 
in our opinion suggest that future improvement of the LCBA equations will be needed to match the genuine weak coupling region.

\newpage
\section{Analyticity of ${\Delta_L(\lambda)}$ in the $\mathfrak{su}(1|1)$ sector at large $\lambda$}
\label{app:analytic}

\begin{theorem}
The solution of the light cone equation Eq.~(\ref{eq:delta11}) for ${\Delta_L(\lambda)}$ at fixed $L$
admits an analytic expansion at large $\lambda$ of the form  
\be
{\Delta_L(\lambda)} = \sum_{n=0}^\infty a_n\, (\lambda^{1/4})^{1-n},
\ee
with a finite radius of convergence. 
\end{theorem}
\proof{
we start again from 
\ba
{\Delta_L(\lambda)} &=& \frac{L}{2} + \sum_{k=1}^L\sqrt{1+\frac{\lambda}{\pi^2}\sin^2\frac{2\pi\,n_k}{{\Delta_L(\lambda)}+L/2}}, \\
\{n_k\} &=& \left\{-\frac{L-1}{2}, \dots, 0, \dots, \frac{L-1}{2}\right\}.
\ea
We set 
\be
w = \frac{2\pi}{{\Delta_L(\lambda)} + L/2},\qquad x^4 = \frac{\pi^2}{\lambda}.
\ee
The equation becomes
\be
\frac{2\pi}{w}-L = 1+2\sum_{k=1}^{\frac{L-1}{2}}\sqrt{1+\frac{1}{x^4}\sin^2(k\,w)},
\ee
or, equivalently:
\be
2\pi\,x^2-(L+1)\,x^2\,w-2\,w\,
\sum_{k=1}^{\frac{L-1}{2}} \sin(k\,w)\sqrt{1+\frac{x^4}{\sin^2(k\,w)}} = 0.
\ee
We scale $z = x/w$ and obtain 
\be
\Phi(z, w) = 0,
\ee
where
\be
\Phi(z, w) = 2\pi\,z^2-(L+1) z^2\, w-2\,
\sum_{k=1}^{\frac{L-1}{2}}\,\frac{\sin(k\,w)}{w}\sqrt{1+z^4\, \frac{w^4}{\sin^2(k\,w)}}
\ee
The equation 
\be
\Phi(z_0, 0) = 0,
\ee
has the solution 
\be
z_0^2 = \frac{1}{\pi} \sum_{k=1}^{\frac{L-1}{2}}\, k = \frac{L^2-1}{8\pi}.
\ee
This corresponds to the asymptotic term
\be
\frac{{\Delta_L(\lambda)}}{L}\sim \frac{1}{\sqrt{2}}\left(1-\frac{1}{L^2}\right)^{1/2}\,\lambda^{1/4}.
\ee
Now, the function $\Phi(z, w)$ is holomorphic in a neighborhood of $(z_0, 0)$ and its partial derivatives are non vanishing at that point
since
\be
\left. \frac{\partial\Phi}{\partial z}\right|_{(z_0, 0)} = 4\,\pi\,z_0,\qquad
\left. \frac{\partial\Phi}{\partial w}\right|_{(z_0, 0)} = -(L+1)\, z_0^2.
\ee
Therefore, by exploiting once again the analytic implicit function theorem~\cite{Gunning},
we conclude that both $z(w)$ and $w(z)$ are holomorphic functions and also that $w$ is an analytic function
of $x = \sqrt{\pi}\lambda^{-1/4}$ in a neighbourhood of $x=0$, which is our thesis.
}

\newpage
\section{Convergence properties of the $1/L$ expansion of the gap equation in the $\mathfrak{su}(1|1)$ sector}
\label{app:plana}

We discuss in some details the convergence properties of the $1/L$ expansion of the gap equation in the $\mathfrak{su}(1|1)$ sector.
The difficult piece is the finite sum appearing in the LCBA
\be
h_L = \sum_{k=1}^{\frac{L-1}{2}} \sqrt{1+\frac{\lambda}{\pi^2}\sin^2\frac{2\pi k}{L\,u}} \equiv \sum_{k=1}^{\frac{L-1}{2}} f\left(\frac{k}{L}\right).
\ee
We study this sum treating $\lambda$ and $\Delta/L$ as fixed parameters and discussing the convergence of the Euler-MacLaurin
summation at large $L$. As a little simplification, we keep the leading term in $\Delta$ and therefore set $\Delta/L = u$, 
with a fixed $u\ge 2$. 

The Euler-MacLaurin summation formula gives 
\be
h_L = \int_0^{\frac{L+1}{2L}} f(x)\,dx-\frac{1}{2}\left[f(0) + f\left(\frac{L+1}{2L}\right)\right] + e_L,
\ee
where
\be
\label{eq:rem}
e_L = \sum_{p\ge 1}\frac{B_{2p}}{(2p)!\,L^{2p-1}} \ f^{(2p-1)}\left(\frac{L+1}{2L}\right).
\ee
This is known to be an asymptotic expansion of the Poincare' type, not necessarily  convergent. 
To investigate the convergence properties of Eq.~(\ref{eq:rem}), we exploit an exact representation of $e_L$ at finite $L$ provided by the 
Abel-Plana formula~\cite{Barton:1981db}
\be
\label{eq:ap}
e_L = -i\int_0^\infty\frac{1}{e^{2\pi\rho}-1}\left[f\left(\frac{L+1+2\,i\,\rho}{2L}\right)- 
f\left(\frac{L+1-2\,i\,\rho}{2L}\right)\right]\,
d\rho
\ee
The advantage of this formula is that it can be analytically continued in the $L$ variable in order to study its analytic structure.
>From the formula and the specific form of $f(z)$, we see that there is a cut extending up to $L\to\infty$ forbidding analyticity. As a check, 
we have evaluated several hundreds of terms in the series
\be
e_L = \sum_{k\ge 0} \frac{c_k}{L^k}.
\ee
By the way, this can be done quite efficiently by expanding the integrand
of the Abel-Plana formula and integrating term by term. We have performed the computation
for generic values of $\lambda$, $u$ as well as for the pair $(\lambda, u(\lambda))$ solving Eq.~(\ref{eq:lo}), one easily always find 
that the successive odd coefficients $d_k = c_{2k+1}$ have the leading behavior
\be
\label{eq:asympt}
|d_k| \sim a\,b^k\,k^c\,k^{d\,k},
\ee
with $d>0$ and suitable $a,b,c$.
The convergence radius of the expansion is therefore confirmed to be zero.

As a toy computation explaining the precise origin of this non-analiticity, one can consider the following simpler integral
having the same analytic structure of the Abel-Plana formula for our problem,
\be
I(z) = \int_0^\infty e^{-\rho}\sqrt{1+\rho\,z}.
\ee
The exact integral can be evaluated and its imaginary part is indeed discontinuous at $arg\,z = \pi$ for any 
radius $|z|$ showing the presence of a cut branching from $z=0$.
If we expand the integrand in powers of $z$ and integrate each term, we obtain the asymptotic expansion
\be
I(z) = \sum_{k\ge 0} c_k \,z^k,\qquad c_k = \binom{\frac{1}{2}}{k}\,\Gamma(k+1).
\ee
Using the expansions at large $k$
\ba
\binom{1/2}{k} &\sim& \frac{2}{\sqrt{\pi}}\sin[\pi(k-1/2)]\,k^{-3/2}, \\
\Gamma(k) &\sim& \sqrt{2\pi}\, k^{k-1/2}\, e^{-k},
\ea
we obtain 
\be
|c_k| \sim \frac{1}{\sqrt{2}}\,e^{-k}\,k^{k-1}, 
\ee
which has the same form as Eq.~(\ref{eq:asympt}).

\newpage
\section{Weak coupling expansion of ${\Delta_L(\lambda)}_{\mathfrak{su}(1|1)}$ in the $L\to \infty$ limit}
\label{app:Linfweak}

The weak coupling expansion of the Eq.~(\ref{eq:lo}) is straightforward and we find
\ba
\lim_{L\to\infty}\frac{{\Delta_L(\lambda)}}{L} &=& u-\frac{1}{2} = \frac{3}{2} +\frac{1}{4}\, \frac{\lambda}{\pi^2} 
-\frac{5}{64}\,\left(\frac{\lambda}{\pi^2}\right)^2 + \frac{5}{128}\,\left(\frac{\lambda}{\pi^2}\right)^3 + \nonumber\\
&& + \frac{4\pi^2-1179}{49152}\,\left(\frac{\lambda}{\pi^2}\right)^4 + \frac{3240-29\,\pi^2}{196608}\,\left(\frac{\lambda}{\pi^2}\right)^5 + {\cal O}(\lambda^6).
\ea
Replacing $u$ in $z_2$ we also obtain 
\be
z_2 = \pi^2\left(-\frac{1}{192}\, \left(\frac{\lambda}{\pi^2}\right)^2 + \frac{19}{3072}\,\left(\frac{\lambda}{\pi^2}\right)^3 + 
\frac{\pi^2-462}{73728}\,\left(\frac{\lambda}{\pi^2}\right)^4 + \dots\right).
\ee
The agreement with our previous results Eq.~(\ref{zzz}) is complete.

\newpage
\section{The expansion of $\mbox{E}(x, -t)$ for $t\to +\infty$}
\label{app:expansion}

\begin{theorem}
The incomplete elliptic integral $\mbox{E}(x, -t)$ with fixed $x>0$, admits the expansion for $t\to + \infty$ 
\be
\mbox{E}(x, -t) = h_0(x)\sqrt{t} + \sum_{n=1}^\infty\frac{h_n(x) + c_n\,\log(16\,t)}{t^{n-\frac{1}{2}}},
\ee
where the first terms of the expansion are 
\ba
\mbox{E}(x, -t) &=& (1-\cos\,x)\sqrt{t} + \\
&& + \frac{1}{4\sqrt{t}}\left(1+\log(16 t) + 2\log\tan\frac{x}{2}\right) + \nonumber \\
&& + \frac{1}{64\,t^{3/2}}\left(3-2\log(16t)-4\log\tan\frac{x}{2}+4\frac{\cos\,x}{\sin^2\,x}\right) + \dots \nonumber 
\ea
and the other are explicitly constructed in the proof.
\end{theorem}
\proof{
First we split
\be
\label{app}
\mbox{E}(x, -t) = \mbox{E}(-t) -\int_x^{\pi/2}\sqrt{1+t\,\sin^2\theta} \, d\theta,
\ee
where $\mbox{E}(-t)$ is the complete elliptic integral of the second kind
\be
\mbox{E}(-t) = \int_0^{\pi/2}\sqrt{1+t\,\sin^2\theta} \, d\theta = \frac{\pi}{2}\,{}_2F_1\left(-\frac 1 2, \frac 1 2 , 1, -t\right).
\ee
Its expansion for large $t$ is non trivial since in the integral the quantity $t\,\sin^2\theta$ is not large when $\theta\to 0$.

\bigskip
The asymptotic expansion can be derived  by using the formula
\ba
{}_2F_1(a, a+m, c, z) &=& \frac{\Gamma(c)(-z)^{-a-m}}{\Gamma(a+m)\Gamma(c-a)}\sum_{n=0}^\infty
\frac{(a)_{n+m}(1-c+a)_{n+m}}{n!(n+m)!} z^{-n}\left[\log(-z) + \right. \nonumber\\
&& \left. + \psi(1+m+n)+\psi(1+n)-\psi(a+m+n)-\psi(c-a-m-n)\right] + \nonumber \\
&& +(-z)^a\frac{\Gamma(c)}{\Gamma(a+m)}\sum_{n=0}^{m-1}\frac{\Gamma(m-n)(a)_n}{n!\Gamma(c-a-n)} z^{-n}
\ea
which is valid for $|\mbox{arg}(-z)|<\pi$, $|z|>1$ and $c-a\not\in\zz$. We are interested in the case $a = -1/2$, $m=1$ and $c=1$ 
that gives
\ba
\lefteqn{\mbox{E}(-t) = \frac{\pi}{2}\,{}_2F_1\left(-\frac{1}{2}, \frac{1}{2}, 1, -t\right) = }&&  \\
&& = \sqrt{t} + \frac{1}{\sqrt{t}}\left[
\frac{1+\log 16t}{4} + \frac{3-2\log 16t}{64 t} + \frac{3(\log 16t-2)}{256 t^2}+\frac{5(133-60\log 16t)}{49152 t^3} + \dots 
\right] \nonumber
\ea
This expansion can be applied for $t>1$.

\bigskip
The second integral in Eq.~(\ref{app}) can be expanded at large $t$ provided $t\,\sin^2\,x > 1$ 
as follows
\ba
\int_x^{\pi/2}\sqrt{1+t\,\sin^2\theta} \, d\theta &=& \sum_{k=0}^\infty \binom{1/2}{k}\,t^{\frac{1}{2}-k}\,\int_x^{\pi/2}\frac{1}{\sin^{2k-1}\theta} d\theta = \nonumber \\
&=& \sqrt{t}\,\cos\,x - \sum_{k=1}^\infty \binom{1/2}{k}\,t^{\frac{1}{2}-k}\,I_k(\cos\,x),
\ea
where 
\be
I_k(a) = \int_0^a\frac{1}{(1-u^2)^k} du
\ee
This integral is elementary and reads
\be
I_k(a) = \frac{\Gamma(k-1/2)}{2\sqrt{\pi}\,\Gamma(k)}\log\frac{1+a}{1-a} + \frac{P_k(a)}{(1-a^2)^{k-1}},
\ee
where the polynomials $P_k(a)$ are defined by 
\ba
P_1(a) &=& 0, \\
P_{k+1}(a) &=& \left(1-\frac{1}{2k}\right) (1-a^2)\,P_k(a) + \frac{a}{2k}.
\ea
The first cases are
\ba
P_2(a) &=& \frac{a}{2}, \\
P_3(a) &=& -\frac{a}{8}(3a^2-5), \\
P_4(a) &=&  \frac{a}{48}(15 a^4-40 a^2+33).
\ea
Collecting these results, we obtain 
\be
\int_x^{\pi/2}\sqrt{1+t\,\sin^2\theta} d\theta = \sqrt{t}\,\cos\,x -\frac{1}{2\sqrt{t}} \log\tan\frac{x}{2} + \frac{1}{16\,t^{3/2}}\left(
\log\tan\frac{x}{2}-\frac{\cos\,x}{\sin^2\,x}\right) + \dots
\ee
Combining our results we prove the thesis.
}

\noindent
Remark: from the proof, we see that the expansion is valid if $t\sin^2\, x>1$. 
For our application we have $x\sim t^{-1/4}$ and the expansion can be applied for large $t$.

\newpage

\vskip 2cm
\FIGURE{\epsfig{file=su11.diff.eps,width=12cm}
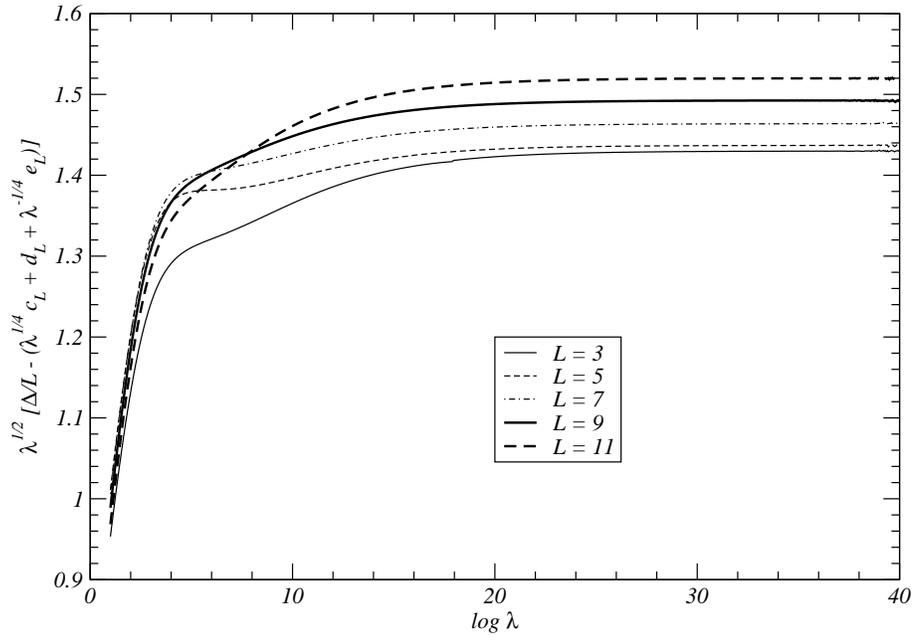
      \caption{Numerical check of the NNLO strong coupling expansion of ${\Delta_L(\lambda)} (\lambda)$ in the $\mathfrak{su}(1|1)$ sector.}
	\label{fig:exp11}}

\vskip 2cm
\FIGURE{\epsfig{file=L.10.scaledp.eps,width=12cm}
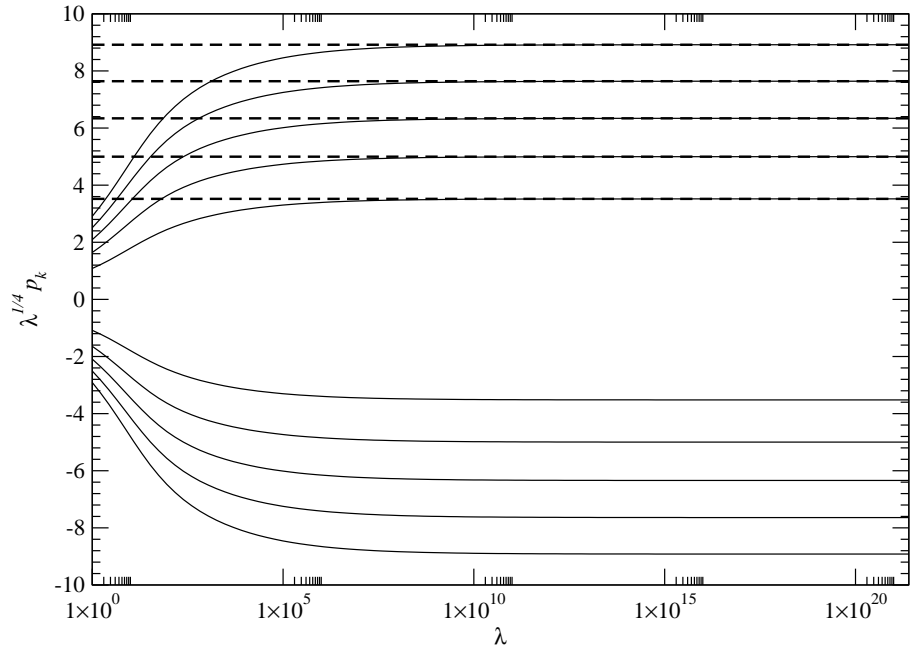
      \caption{Scaled Bethe momenta for the AF state in the $\mathfrak{su}(2)$ sector. Here $L=10$.}
	\label{fig:su2sol}}

\vskip 2cm
\FIGURE{\epsfig{file=alpha.L.150.eps,width=12cm}
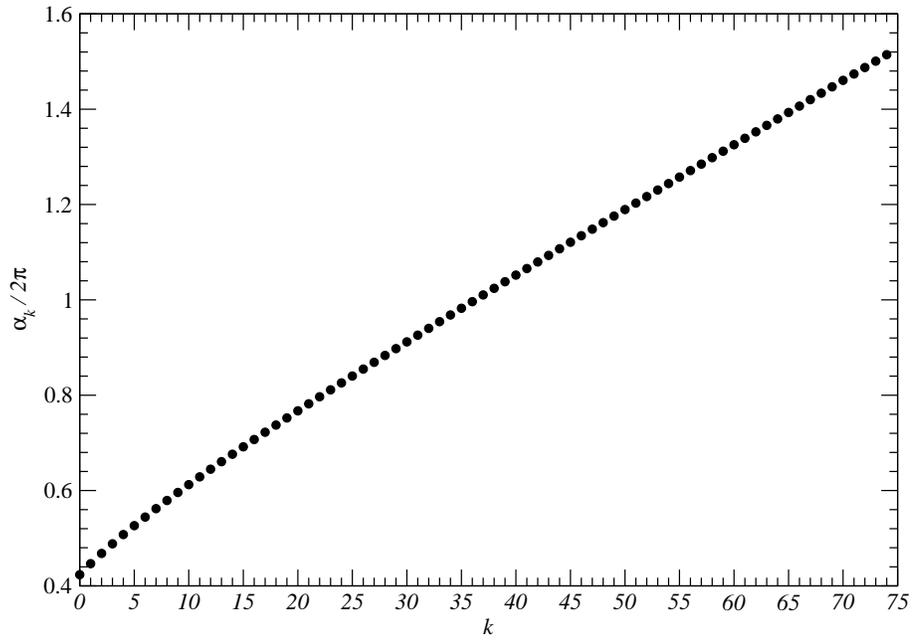
      \caption{Distribution of $\alpha_k$ for $L=150$.}
	\label{fig:alpha}}

\end{document}